\documentclass[aps,prd,preprint,groupedaddress,nofootinbib]{revtex4}

\usepackage{graphicx}
\usepackage{epsfig}
\usepackage{amsmath}
\usepackage{amsfonts,amssymb}
\usepackage{color}

  \usepackage{slashed}
  \usepackage{url}
\usepackage{bm}
\usepackage{verbatim}
\usepackage{float}

\usepackage{mathtools}

\begin{document}


\title{Warping effects in strongly perturbed metrics}


\author{Marco Frasca}
\author{Riccardo Maria Liberati}
\author{Massimiliano Rossi}
\affiliation{MBDA Italia S.p.A., Via Monte Flavio, 45,\\
00131, Rome (Italy)}


\date{\today}

\begin{abstract}
A technique devised some years ago permits to study a theory in a regime of strong perturbations. This translates into a gradient expansion that, at the leading order, can recover the BKL solution in general relativity. We solve exactly the leading order Einstein equations in a spherical symmetric case, assuming a Schwarzschild metric under the effect of a time-dependent perturbation, and we show that the 4-velocity in such a case is multiplied by an exponential warp factor when the perturbation is properly applied. This factor is always greater than one. We will give a closed form solution of this factor for a simple case. Some numerical examples are also given.
\end{abstract}


\maketitle



\section{Introduction}

The study of Einstein equations in certain regimes is often reduced to solve them numerically \cite{teul99}. The reason is that they form a set of nonlinear PDEs that are generally difficult to handle with analytical tools for most interesting situations. Often, the reason relies on the fact that no small parameter can be found to apply standard perturbation techniques while analytical solutions are very rare and difficult to find. Some years ago, one of us (M.F.) proposed an approach based on earlier works in strongly perturbed systems \cite{fra06}. It was shown that, under a strong perturbation in the formal limit running to infinity, the leading order is obtained by neglecting the gradient terms in the Einstein equations. The leading order of this perturbation series was firstly proposed by Belinsky, Kalathnikov, Lifshitz for their famous BKL conjecture \cite{kl70,bkl70,bkl82}, as is known today. 

Some decades ago, Alcubierre proposed a solution of the Einstein equations \cite{alcu94} that describes an observer moving with an unbounded velocity provided the condition of positivity of the energy is violated. A recent paper \cite{Santos-Pereira:2020puq} (see also Refs. therein) yields a short recount about Alcubierre metric and its interaction with dust. Indeed, any kind of pathology has emerged about it and the difficulties arise from the fact that this is an engineered metric that is imposed on the Einstein equations. It would be desirable to have a metric like this one emerging as a solution of the Einstein equations and conserving the positivity of the energy. A recent proposal goes in such a direction \cite{Lentz:2020euv}. This is possible by introducing a hyperbolic shift vector potential and the author shows how this can emerge from a plasma.

In this paper we will show how a warp factor for the velocity can emerge when a strong perturbation is applied to a spherical symmetric metric. So, any Eulerian observer will get its velocity expanded when such perturbation is acting. This extends and complete our preceding work \cite{fra06}. We will get the exact solution of the leading perturbation equations and we will show how an exponential factor can emerge that is systematically greater than one.
We emphasize that we are doing perturbation theory in a limit where the a perturbation, applied to a given gravitational field, is taken much greater of the unperturbed situation. This is the opposite limit to standard small perturbation theory and is based on the technique devised in \cite{fra06}.

The paper is so structured. In Sec.\ref{sec1}, we will introduce the technique to treat strongly perturbed systems. In Sec.\ref{sec2}, we apply this to the Einstein equations for a spherical symmetry metric with a time-dependent perturbation. In Sec.\ref{sec3}, we solve the leading order perturbation equations. In Sec.\ref{sec4}, we yield the geodesic equations. In Sec.\ref{sec5}, we show how the expansion factor enters into the velocity providing some examples and an analytic solution. In Sec.\ref{sec6}, conclusions are presented.

\section{Strong perturbations and gradient expansion}
\label{sec1}

For our computations in general relativity, we need to study the case of a strong perturbation on a given metric that we will choose to be the Schwarzschild one. In order to prove that a gradient expansion indeed represents a strong perturbation theory, we will study the following non-linear equation as a toy model for the Einstein equations\footnote{We just point out that this model can represent Einstein equations in 1+1 dimensions \cite{teitel, jack, DHoker:1982wmk}.} (here and in the following we assume $c=1$). 
%
\begin{equation}
    -\Box\phi+\lambda V'(\phi)=0
\end{equation}
being $\Box=\nabla^2-\partial^2_t$ the wave operator (here and in the following $c=1$), $\phi$ a scalar field and $V(\phi)$ its self-interaction with a coupling $\lambda$. For 2D Einstein equations, this would be a Liouville equation \cite{teitel, jack, DHoker:1982wmk}. We would like to do perturbation theory in the formal limit of $\lambda\rightarrow\infty$. This ends up to obtain a non-trivial series in $1/\lambda$. We can accomplish our aim by rescaling the time variable \cite{Frasca:1998ch}. We take $t\rightarrow\sqrt{\lambda}t$ and the equation above becomes
\begin{equation}
\label{eq:scal}
    -\nabla^2\phi+\lambda\partial^2_t\phi+\lambda V'(\phi)=0.
\end{equation}
Then, we take
\begin{equation}
    \phi=\phi_0+\frac{1}{\lambda}\phi_1+\frac{1}{\lambda^2}\phi_2+\ldots
\end{equation}
and substitute this into eq.(\ref{eq:scal}). This gives the set of perturbative equations
\begin{eqnarray}
    \partial^2_t\phi_0&=&-V'(\phi_0) \nonumber \\
    \partial^2_t\phi_1&=&-V''(\phi_0)\phi_1+\nabla^2\phi_0 \nonumber
    \\
    \partial_t^2\phi_2&=&-V''(\phi_0)\phi_2-\frac{1}{2}V'''(\phi_0)\phi_1^2+\nabla^2\phi_1 \nonumber \\
    &\vdots&.
\end{eqnarray}
We see that we have obtained a set of non-trivial equations that define the perturbation series in the formal limit $\lambda\rightarrow\infty$. This approach can be applied, exactly in this way, to the Einstein equations. This also shows how consistent was the original BKL approach in \cite{kl70,bkl70,bkl82}. Indeed, we have obtained a gradient expansion. 

In order to see how this technique applies to Einstein equations, we write them down in the Arnowitt-Deser-Misner (ADM) formalism as \cite{teul99}
\begin{eqnarray}
    \partial_t\gamma_{ij}-\beta^l\partial_l\gamma_{ij} &=& \gamma_{lj}\partial_i\beta^l+
    \gamma_{il}\partial_j\beta^l-2\alpha K_{ij} \\ \nonumber
    \partial_tK_{ij}-\beta^l\partial_lK_{ij} &=& K_{il}\partial_j\beta^l+K_{jl}\partial_i\beta^l
    -2\alpha K_{il}K_j^l+\alpha K K_{ij} \\ \nonumber
    & &-\frac{1}{2}\alpha\gamma^{lm}\left\{\partial_l\partial_m\gamma_{ij}+\partial_i\partial_j\gamma_{lm}
    -\partial_i\partial_l\gamma_{mj}-\partial_j\partial_l\gamma_{mi}\right. \\ \nonumber
    & & +\gamma^{np}\left[(\partial_i\gamma_{jn}+\partial_j\gamma_{in}
    -\partial_n\gamma_{ij})\partial_l\gamma_{mp}\right. \\ \nonumber
    & & +\partial_l\gamma_{in}\partial_p\gamma_{jm}-\partial_l\gamma_{in}\partial_m\gamma_{jp}\left.\right] \\ \nonumber
    & &-\frac{1}{2}\gamma^{np}\left[(\partial_i\gamma_{jn}+\partial_j\gamma_{in}
    -\partial_n\gamma_{ij})\partial_p\gamma_{lm}+
    \partial_i\gamma_{ln}\partial_j\gamma_{mp}\left.\right]\right\} \\ \nonumber
    & &-\partial_i\partial_j\alpha+\frac{1}{2}\gamma^{lm}(\partial_i\gamma_{jm}+\partial_j\gamma_{im}
    -\partial_m\gamma_{ij})\partial_l\alpha \nonumber \\
    & &+\alpha\left[-8\pi GT_{ij}+4\pi G\gamma_{ij}(T-\rho)\right]
\end{eqnarray}
where the energy-matter tensor $T_{\mu\nu}$ is given with the density $\rho$, 
for a metric
\begin{equation}
    ds^2 = (-\alpha^2 + \gamma_{ij}\beta^i\beta^j)dt^2+2\beta^idx_idt+\gamma_{ij}dx^idx^j
\end{equation}
being $\alpha$ the lapse function, $\beta^i$ the shift vector, $\gamma_{ij}$ the spatial part of the metric and $K_{ij}$ the extrinsic curvature. For our aims, we do not care about constraint equations that are just essential for numerical computations. This set is amenable to the same treatment we applied to the preceding example. The procedure is identical, we introduce an ordering parameter $\lambda$ that we will set to 1 to the end of computation. Then, we consider the perturbation series defined by
\begin{eqnarray}
    \tau &=& \sqrt{\lambda}t,\\ \nonumber
	  K_{ij} &=& \sqrt{\lambda}\left(K_{ij}^{(0)}+\frac{1}{\lambda}K_{ij}^{(1)}+\frac{1}{\lambda^2}K_{ij}^{(2)}
	  +\ldots\right), \\ \nonumber
	  \gamma_{ij} &=& \gamma_{ij}^{(0)}+\frac{1}{\lambda}\gamma_{ij}^{(1)}+\frac{1}{\lambda^2}\gamma_{ij}^{(2)}
	  +\ldots, \\ \nonumber
	  \alpha &=& \alpha_0 + \frac{1}{\lambda}\alpha_1+\frac{1}{\lambda^2}\alpha_2+\ldots.
\end{eqnarray}
Our gauge choice is to set the shift vector $\beta_i=0$. This approach is completely dual to the standard computation for weak gravitational fields but implies that the perturbation is taken formally to go to infinity. This represents a situation where the perturbation overcomes the intensity of the gravitational field where is applied. Typical situations where this technique could apply are black hole collisions where, currently, only numerical computations or analytical techniques, working given certain approximations, are available \cite{Soffel:2019aoq}. Therefore, we get the non-trivial set of equations (we have set $\lambda=1$)
\begin{eqnarray}
\label{eq:Set0}
    \partial_t\gamma_{ij}^{(0)} &=& -2\alpha_0K_{ij}^{(0)} \\ \nonumber
    \partial_t\gamma_{ij}^{(1)} &=& -2\alpha_1K_{ij}^{(0)}-2\alpha_0K_{ij}^{(1)} \\ \nonumber
    &\vdots& \\ \nonumber
    \partial_t K_{ij}^{(0)} &=& -2\alpha_0K_{il}^{(0)}K_{j}^{l(0)} + \alpha_0K^{(0)}K_{ij}^{(0)} \\ \nonumber
    \partial_t K_{ij}^{(1)} &=& -2\alpha_1K_{il}^{(0)}K_{j}^{l(0)}-2\alpha_0K_{il}^{(1)}K_{j}^{l(0)}
                                     -2\alpha_0K_{il}^{(0)}K_{j}^{l(1)} \\ \nonumber
                                 & & + \alpha_1K^{(0)}K_{ij}^{(0)} + \alpha_0K^{(1)}K_{ij}^{(0)}   
                                     + \alpha_0K^{(0)}K_{ij}^{(1)} \\ \nonumber
    & &-\frac{1}{2}\alpha_0\gamma^{lm(0)}\left\{\partial_l\partial_m\gamma_{ij}^{(0)}
    +\partial_i\partial_j\gamma_{lm}^{(0)}
    -\partial_i\partial_l\gamma_{mj}^{(0)}-\partial_j\partial_l\gamma_{mi}^{(0)}\right. \\ \nonumber
    & & +\gamma^{np(0)}\left[(\partial_i\gamma_{jn}^{(0)}+\partial_j\gamma_{in}^{(0)}
    -\partial_n\gamma_{ij}^{(0)})\partial_l\gamma_{mp}^{(0)}\right. \\ \nonumber
    & & +\partial_l\gamma_{in}^{(0)}\partial_p\gamma_{jm}^{(0)}
    -\partial_l\gamma_{in}^{(0)}\partial_m\gamma_{jp}^{(0)}\left.\right] \\ \nonumber
    & &-\frac{1}{2}\gamma^{np(0)}\left[(\partial_i\gamma_{jn}^{(0)}+\partial_j\gamma_{in}^{(0)}
    -\partial_n\gamma_{ij}^{(0)})\partial_p\gamma_{lm}^{(0)}+
    \partial_i\gamma_{ln}^{(0)}\partial_j\gamma_{mp}^{(0)}\left.\right]\right\} \\ \nonumber
    & &-\partial_i\partial_j\alpha_0+\frac{1}{2}\gamma^{lm(0)}(\partial_i\gamma_{jm}^{(0)}+\partial_j\gamma_{im}^{(0)}
    -\partial_m\gamma_{ij}^{(0)})\partial_l\alpha_0 \\ \nonumber
    & &+\alpha^{(0)}\left[-8\pi GT_{ij}+4\pi G\gamma_{ij}^{(0)}(T-\rho)\right] \nonumber \\
    &\vdots&
\end{eqnarray}
where one sees that the energy-matter tensor contributes to the next-to-leading order. We realize from 
these equations that the gradient terms, that is components of the metric that are varying spatially, are moved to the next-to-leading order. We will apply them in the following in the spherical symmetric case, assuming the Schwarzschild metric as the unperturbed solution. 
Here and in the following, we avoid to show explicitly the energy-matter tensor as our perturbation series moves its contribution to the next-to-leading order. This implies that, at the leading order, an approximation for the energy-matter configuration can be taken to be that in absence of the gravitational field. This is consistent with our approach for strongly perturbed metrics.

Anyway, in order to have an idea of the main concept underlying this approximation scheme, let us consider the Reissner-N\"ordstrom metric of a charged black hole. This will be given by
\begin{equation}
    ds^2=-\left(1-\frac{r_g}{r}+\frac{r_Q^2}{r^2}\right)dt^2
    +\left(1-\frac{r_g}{r}+\frac{r_Q^2}{r^2}\right)^{-1}dr^2
    +r^2d\theta^2+r^2\sin^2\theta d\phi^2.
\end{equation}
being $r_g=2GM$ the Schwarzschild radius and $r_q^2=GQ^2/4\pi\epsilon_0$ the scale introduced by the black hole charge $Q$ with $1/4\pi\epsilon_0$ the Coulomb constant. This is an exact solution of the Einstein-Maxwell equations. In our case, we assume that the electric field overcomes largely the gravitational contribution, that is $r_Q\gg r_g$. This appears formally as a large perturbation on a Schwarzschild black hole and the approximate metric will be
\begin{equation}
    ds^2=-\left(1+\frac{r_Q^2}{r^2}\right)dt^2
    +\left(1+\frac{r_Q^2}{r^2}\right)^{-1}dr^2
    +r^2d\theta^2+r^2\sin^2\theta d\phi^2+O(r_g/r).
\end{equation}
This should be compared with the opposite dual limit $r_Q/r_g\ll 1$ that yields
\begin{equation}
    ds^2=-\left(1-\frac{r_g}{r}\right)dt^2
    +\left(1-\frac{r_g}{r}\right)^{-1}dr^2
    +r^2d\theta^2+r^2\sin^2\theta d\phi^2+O(r_Q^2/r^2).
\end{equation}

\section{Strongly perturbed spherical symmetry metric}
\label{sec2}


We are assuming a spherical symmetry metric in ADM formalism given by
\begin{equation}
    ds^2=-\alpha^2 dt^2+\gamma_{rr}dr^2+\gamma_{\theta\theta}d\theta^2+
    \gamma_{\phi\phi}d\phi^2.
\end{equation}
This implies a specific choice of the gauge where all the components of the shift vector, normally named $\beta_i$, are taken to be zero. Then, the perturbation $\alpha_1$ is just applied to the lapse function as follows \cite{teul99}
\begin{equation}
    \alpha^2=\alpha_0^2+\alpha_1.
\end{equation}
Then, we specialize the set of eq.(\ref{eq:Set0}) to this case.
Assuming as unperturbed solution the Schwarzschild one, the exterior solution is given by (again, $r_g=2GM$ is the Schwarzschild radius)
\begin{eqnarray}
    \alpha_0^2 &=& \left(1-\frac{r_g}{r}\right), \\ \nonumber
	\gamma_{11}^{(0)} &=& \frac{1}{1-\frac{r_g}{r}}, \\ \nonumber
	\gamma_{22}^{(0)} = r^2, & & \gamma_{33}^{(0)} = r^2\sin^2\theta,
\end{eqnarray}
and the interior solution is
\begin{eqnarray}
    \alpha_0^2 &=& \frac{1}{4} \left( 3 \sqrt{1-\frac {r_g}{r_s}}-\sqrt{1-\frac{r^2 r_g}{r_s^3}} \right)^2, \\ \nonumber
	\gamma_{11}^{(0)} &=& \left( 1-\frac{r^2 r_g}{r_s^3} \right)^{-1}, \\ \nonumber
	\gamma_{22}^{(0)} = r^2 & & \gamma_{33}^{(0)} = r^2\sin^2\theta,
\end{eqnarray}
being $r_s$ is the value of the $r$-coordinate at the body's surface.
It easy to see that both metrics are the same at the sphere surface for $r=r_s$ granting continuity.
We also have, with our gauge's choice $\beta_i=0$, the general formula
\begin{equation}
\label{eq:Kgamma}
    K_{ij}=-\frac{1}{2\alpha}\partial_t\gamma_{ij}.
\end{equation}
In our case is
\begin{equation}
    \alpha^2=\alpha_0^2+\alpha_1=\alpha_0^2+A f(r,t),
\end{equation}
being $A$ the amplitude of the perturbation. We emphasize that the perturbations we are going to consider are time-dependent. This yields
\begin{equation}
    \partial_t\gamma_{ij}^{(1)} = \alpha_1\frac{1}{\alpha_0}\partial_t\gamma_{ij}^{(0)}-2\alpha_0K_{ij}^{(1)},
\end{equation}
that reduces to
\begin{equation}
    \partial_t\gamma_{ij}^{(1)} =-2\alpha_0K_{ij}^{(1)},
\end{equation}
as $\gamma_{ij}^{(0)}$ does not depend on time variable. Now, one has
\begin{eqnarray}
 \partial_t K_{ij}^{(1)} &=& -2\alpha_1K_{il}^{(0)}K_{j}^{l(0)}-2\alpha_0K_{il}^{(1)}K_{j}^{l(0)}
 -2\alpha_0K_{il}^{(0)}K_{j}^{l(1)} \\ \nonumber
& & + \alpha_1K^{(0)}K_{ij}^{(0)} + \alpha_0K^{(1)}K_{ij}^{(0)}   
    + \alpha_0K^{(0)}K_{ij}^{(1)} \\ \nonumber
    & &+ \alpha_1K^{(0)}K_{ij}^{(0)} + \alpha_0K^{(1)}K_{ij}^{(0)}   
    + \alpha_0K^{(0)}K_{ij}^{(1)} \\ \nonumber
    & &-\frac{1}{2}\alpha_0\gamma^{lm(0)}\left\{\partial_l\partial_m\gamma_{ij}^{(0)}
    +\partial_i\partial_j\gamma_{lm}^{(0)}
    -\partial_i\partial_l\gamma_{mj}^{(0)}-\partial_j\partial_l\gamma_{mi}^{(0)}\right. \\ \nonumber
    & & +\gamma^{np(0)}\left[(\partial_i\gamma_{jn}^{(0)}+\partial_j\gamma_{in}^{(0)}
    -\partial_n\gamma_{ij}^{(0)})\partial_l\gamma_{mp}^{(0)}\right. \\ \nonumber
    & & +\partial_l\gamma_{in}^{(0)}\partial_p\gamma_{jm}^{(0)}
    -\partial_l\gamma_{in}^{(0)}\partial_m\gamma_{jp}^{(0)}\left.\right] \\ \nonumber
    & &-\frac{1}{2}\gamma^{np(0)}\left[(\partial_i\gamma_{jn}^{(0)}+\partial_j\gamma_{in}^{(0)}
    -\partial_n\gamma_{ij}^{(0)})\partial_p\gamma_{lm}^{(0)}+
    \partial_i\gamma_{ln}^{(0)}\partial_j\gamma_{mp}^{(0)}\left.\right]\right\} \\ \nonumber
    & &-\partial_i\partial_j\alpha_0+\frac{1}{2}\gamma^{lm(0)}(\partial_i\gamma_{jm}^{(0)}+\partial_j\gamma_{im}^{(0)}
    -\partial_m\gamma_{ij}^{(0)})\partial_l\alpha_0 \\ \nonumber
    &\vdots&.
\end{eqnarray}
This set of equations, written in this way, are too difficult to manage. As we will see below, we can restate them to find an exact leading order solution.

It is correct to ask why the Birkhoff theorem does not apply in our case. The reason is that the problems we are treating are similar to the one of the ringdown of a Schwarzschild black-hole where a strong perturbation, due to the collision between two black holes, modifies the metric making it varying in time, after coalescence, until the oscillations are damped out and the spherical symmetry is recovered \cite{Merritt:2004gc}, in agreement with Birkhoff theorem. It should be said that such problems are better managed in the Kerr metric but we do not consider rotations to avoid too much computations cluttering formulas.

\section{Solving perturbation equations}
\label{sec3}

In order to have more manageable equations, let us start from the following rewriting of the ADM equations of motion in exact form. We will get (as already said, our gauge is $\beta_i=0$)
\begin{eqnarray}
    \partial_t\gamma_{ij}&=&-2\alpha K_{ij} \nonumber \\
    \partial_t K_{ij}&=&\alpha\left[R_{ij}-2K_{il}K_j^l+KK_{ij}\right]-\partial_i\partial_j\alpha.
\end{eqnarray}
The Ricci tensor $R_{ij}$ refers to the $\gamma_{ij}$ and all Latin indexes run from 1 to 3. We can exploit these equations for the diagonal elements to obtain
\begin{eqnarray}
    \partial_t\gamma_{11}&=&-2\alpha K_{11}, \nonumber \\
    \partial_t\gamma_{22}&=&-2\alpha K_{22}, \nonumber \\
    \partial_t\gamma_{33}&=&-2\alpha K_{33}, \nonumber \\
    \partial_t K_{11}&=&\alpha\left[R_{11}-2K_{1l}K_1^l+KK_{11}\right]-\partial_1^2\alpha, \nonumber \\
    \partial_t K_{22}&=&\alpha\left[R_{22}-2K_{2l}K_2^l+KK_{22}\right]-\partial_2^2\alpha, \nonumber \\
    \partial_t K_{33}&=&\alpha\left[R_{33}-2K_{3l}K_3^l+KK_{33}\right]-\partial_3^2\alpha.
\end{eqnarray}
We notice that $K_i^l=\gamma^{kl}K_{ik}$ and $K=\gamma^{kl}K_{kl}$.
We expect that off-diagonal terms should be perturbatively negligible and so, we neglect them here in view of a gradient expansion. Indeed, for $i\ne j$, we will have
\begin{eqnarray}
    K_{ij} &=& -\frac{\partial_t\gamma_{ij}}{2\alpha} \nonumber \\
    \partial_t K_{ij}&=&\alpha\left[R_{ij}-2K_{il}\gamma^{lk}K_{jk}+\gamma^{kl}K_{kl}K_{ij}\right]-\partial_i\partial_j\alpha.
\end{eqnarray}
This will give
\begin{equation}
\label{eq:off-d}
    -\partial_t\left(\frac{\partial_t\gamma_{ij}}{2\alpha}\right)
    =\alpha\left[R_{ij}+\frac{1}{2\alpha^2}\partial_t\gamma_{il}\gamma^{lk}\partial_t\gamma_{jk}+\frac{1}{4\alpha^2}\gamma^{kl}\partial_t\gamma_{kl}\partial_t\gamma_{ij}\right]-\partial_i\partial_j\alpha.
\end{equation}
In a gradient expansion, where we neglect both $R_{ij}$ and $\partial_i\partial_j\alpha$ as we will show below, at the leading order the off-diagonal terms will remain 0 if they were zero initially because this is a solution of eq.(\ref{eq:off-d}) for $i\ne j$.
Therefore,
\begin{eqnarray}
    \partial_t\gamma_{11}&=&-2\alpha K_{11},  \\
    \partial_t\gamma_{22}&=&-2\alpha K_{22}, \nonumber \\
    \partial_t\gamma_{33}&=&-2\alpha K_{33}, \nonumber \\
    \partial_t K_{11}&=&\alpha\left[R_{11}-\gamma^{11}K_{11}^2+(\gamma^{22}K_{22}+\gamma^{33}K_{33})K_{11}\right]-\partial_1^2\alpha, \nonumber \\
    \partial_t K_{22}&=&\alpha\left[R_{22}-\gamma^{22}K_{22}^2+(\gamma^{11}K_{11}+\gamma^{33}K_{33})K_{22}\right]-\partial_2^2\alpha, \nonumber \\
    \partial_t K_{33}&=&\alpha\left[R_{33}-\gamma^{33}K_{33}^2+(\gamma^{11}K_{11}+\gamma^{22}K_{22})K_{33}\right]-\partial_3^2\alpha. \nonumber
\end{eqnarray}
These equation can be stated in a single set of equations for the $\gamma$s as
\begin{eqnarray}
    \partial_t^2\gamma_{11}&=&-2\dot\alpha K_{11}-2\alpha\dot K_{11}= \nonumber \\
    &&\frac{\dot\alpha}{\alpha}\dot\gamma_{11}-2\alpha^2\left[R_{11}
    -\gamma^{11}\frac{1}{4\alpha^2}(\dot\gamma_{11})^2
    +\frac{1}{4\alpha^2}\gamma^{22}\dot\gamma_{22}\dot\gamma_{11}
     +\frac{1}{4\alpha^2}\gamma^{33}\dot\gamma_{33}\dot\gamma_{11}
    \right]+2\alpha\partial_1^2\alpha \nonumber \\
    \partial_t^2\gamma_{22}&=&-2\dot\alpha K_{22}-2\alpha\dot K_{22}=
    \nonumber \\
    &&\frac{\dot\alpha}{\alpha}\dot\gamma_{22}-2\alpha^2\left[R_{22}
    -\gamma^{22}\frac{1}{4\alpha^2}(\dot\gamma_{22})^2
    +\frac{1}{4\alpha^2}\gamma^{11}\dot\gamma_{11}\dot\gamma_{22}
     +\frac{1}{4\alpha^2}\gamma^{33}\dot\gamma_{33}\dot\gamma_{22}
    \right]+2\alpha\partial_2^2\alpha \nonumber \\
    \partial_t^2\gamma_{33}&=&-2\dot\alpha K_{33}-2\alpha\dot K_{33}=
    \\
    &&\frac{\dot\alpha}{\alpha}\dot\gamma_{33}-2\alpha^2\left[R_{33}
    -\gamma^{33}\frac{1}{4\alpha^2}(\dot\gamma_{33})^2
    +\frac{1}{4\alpha^2}\gamma^{11}\dot\gamma_{11}\dot\gamma_{33}
     +\frac{1}{4\alpha^2}\gamma^{22}\dot\gamma_{22}\dot\gamma_{33}
    \right]+2\alpha\partial_3^2\alpha. \nonumber
\end{eqnarray}
and so on for the other components. As said into Sec.\ref{sec2}, this set of equations can be solved perturbatively by the change of variable $\tau=\sqrt{\lambda}t$ being $\lambda$ just an ordering parameter that we will set to 1 to the end of computations. 
This means that we can neglect spatial gradients at the leading order, yielding
\begin{eqnarray}
    \partial_\tau^2\gamma_{11}&=&
    \frac{\dot\alpha}{\alpha}\dot\gamma_{11}
    +\frac{1}{2}\gamma^{11}(\dot\gamma_{11})^2
    -\frac{1}{2}\gamma^{22}\dot\gamma_{22}\dot\gamma_{11}
     -\frac{1}{2}\gamma^{33}\dot\gamma_{33}\dot\gamma_{11}= \nonumber \\
    &&\frac{\dot\alpha}{\alpha}\dot\gamma_{11}
    +\frac{1}{2}(\gamma_{11})^{-1}(\dot\gamma_{11})^2
    -\frac{1}{2}(\gamma_{22})^{-1}\dot\gamma_{22}\dot\gamma_{11}
     -\frac{1}{2}(\gamma_{33})^{-1}\dot\gamma_{33}\dot\gamma_{11}
\end{eqnarray}
This can be rewritten as
\begin{equation}
   \partial_\tau^2\gamma_{11}= \dot\gamma_{11}\frac{d}{d\tau}\left[\ln\alpha+\frac{1}{2}\ln\left(\frac{\gamma_{11}}{\gamma_{22}\gamma_{33}}\right)\right]
\end{equation}
Then,
\begin{equation}
   \partial_\tau\ln\dot\gamma_{11}=\frac{d}{d\tau}\left[\ln\alpha+\frac{1}{2}\ln\left(\frac{\gamma_{11}}{\gamma_{22}\gamma_{33}}\right)\right]
\end{equation}
and finally
\begin{equation}
   \ln\dot\gamma_{11}=\left[\ln\left(r_k\frac{\alpha}{\alpha_0}\right)+\frac{1}{2}\ln\left(\frac{\gamma_{11}}{\gamma_{22}\gamma_{33}}\right)\right]
\end{equation}
where we have properly fixed the integration constant in such a way that, in absence of perturbation, the contribution from $\alpha$ disappears while dimensions are kept with the constant $r_k=r_g$ for the exterior solution and $r_k=r_s$ for the interior solution. This gives the following set of differential equations
\begin{eqnarray}
    \dot\gamma_{11}&=&r_k\frac{\alpha}{\alpha_0}\sqrt{\frac{\gamma_{11}}{\gamma_{22}\gamma_{33}}}=r_k\frac{\alpha}{\alpha_0}\gamma_{11}\gamma^{-\frac{1}{2}} \nonumber \\
    \dot\gamma_{22}&=&r_k\frac{\alpha}{\alpha_0}\sqrt{\frac{\gamma_{22}}{\gamma_{11}\gamma_{33}}}=r_k\frac{\alpha}{\alpha_0}\gamma_{22}\gamma^{-\frac{1}{2}} \nonumber \\
    \dot\gamma_{33}&=&r_k\frac{\alpha}{\alpha_0}\sqrt{\frac{\gamma_{33}}{\gamma_{11}\gamma_{22}}}=r_k\frac{\alpha}{\alpha_0}\gamma_{33}\gamma^{-\frac{1}{2}}
\end{eqnarray}
This set can be solved exactly by multiplying in the following way
\begin{eqnarray}
    \dot\gamma_{11}\gamma_{22}\gamma_{33}&=&r_k\frac{\alpha}{\alpha_0}\gamma^{\frac{1}{2}} \nonumber \\
    \dot\gamma_{22}\gamma_{11}\gamma_{33}&=&r_k\frac{\alpha}{\alpha_0}\gamma^{\frac{1}{2}} \nonumber \\
    \dot\gamma_{33}\gamma_{11}\gamma_{22}&=&r_k\frac{\alpha}{\alpha_0}\gamma^{\frac{1}{2}}
\end{eqnarray}
and summing up the three equations obtained in this way giving
\begin{equation}
    \dot\gamma=3r_k\frac{\alpha}{\alpha_0}\gamma^{\frac{1}{2}}
\end{equation}
that has as a solution
\begin{equation}
    \gamma(t)=\left[\frac{3}{2}r_k\alpha_0^{-1}\int_0^t\alpha(t')dt'+\sqrt{\gamma(0)}\right]^2
\end{equation}
and, e.g. one has
\begin{equation}
    \gamma(0)=|\gamma_{11}^{(0)}\gamma_{22}^{(0)}\gamma_{33}^{(0)}|=\frac{r^4\sin^2\theta}{1-\frac{r_g}{r}}.
\end{equation}
for the exterior solution. This yields the set of equations
\begin{eqnarray}
    \dot\gamma_{11}&=&\frac{r_k\alpha_0^{-1}\alpha}{\frac{3}{2}r_k\alpha_0^{-1}\int_0^t\alpha(t')dt'+\sqrt{\gamma(0)}}\gamma_{11} \nonumber \\
    \dot\gamma_{22}&=&\frac{r_k\alpha_0^{-1}\alpha}{\frac{3}{2}r_k\alpha_0^{-1}\int_0^t\alpha(t')dt'+\sqrt{\gamma(0)}}\gamma_{22} \nonumber \\
    \dot\gamma_{33}&=&\frac{r_k\alpha_0^{-1}\alpha}{\frac{3}{2}r_k\alpha_0^{-1}\int_0^t\alpha(t')dt'+\sqrt{\gamma(0)}}\gamma_{33}.
\end{eqnarray}
These can be solved exactly by
\begin{eqnarray}
\label{eq:gammas}
    \gamma_{11}(t)&=&\exp\left[r_k\alpha_0^{-1}\int_0^tdt''\frac{\alpha(t'')}{\frac{3}{2}r_k\alpha_0^{-1}\int_0^{t''}\alpha(t')dt'+\sqrt{\gamma(0)}}\right]\gamma_{11}^{(0)} \nonumber \\
    \gamma_{22}(t)&=&\exp\left[r_k\alpha_0^{-1}\int_0^tdt''\frac{\alpha(t'')}{\frac{3}{2}r_k\alpha_0^{-1}\int_0^{t''}\alpha(t')dt'+\sqrt{\gamma(0)}}\right]\gamma_{22}^{(0)} \nonumber \\
    \gamma_{33}(t)&=&\exp\left[r_k\alpha_0^{-1}\int_0^tdt''\frac{\alpha(t'')}{\frac{3}{2}r_k\alpha_0^{-1}\int_0^{t''}\alpha(t')dt'+\sqrt{\gamma(0)}}\right]\gamma_{33}^{(0)}.
\end{eqnarray}
We can derive the volume expansion from the equation \cite{alcu94}
\begin{equation}
    \Theta = -\alpha{\rm Tr}K=-\alpha\gamma^{ij}K_{ij}=\frac{1}{2}\gamma^{ij}
    {\dot\gamma_{ij}}
\end{equation}
and $K_{ij}$ are given by eq.(\ref{eq:Kgamma}). Then,
\begin{equation}
    \Theta=\frac{3}{2}\frac{r_k\alpha_0^{-1}\alpha}{\frac{3}{2}r_k\alpha_0^{-1}\int_0^t\alpha(t')dt'+\sqrt{\gamma(0)}}
    \exp\left[r_k\alpha_0^{-1}\int_0^tdt''\frac{\alpha(t'')}{\frac{3}{2}r_k\alpha_0^{-1}\int_0^{t''}\alpha(t')dt'+\sqrt{\gamma(0)}}\right].
\end{equation}
Here we can see the first appearance of the expansion (warp) factor given by
\begin{equation}
    U(r,\theta,t)=\exp\left[r_k\alpha_0^{-1}\int_0^tdt''\frac{\alpha(t'')}{\frac{3}{2}r_k\alpha_0^{-1}\int_0^{t''}\alpha(t')dt'+\sqrt{\gamma(0)}}\right].
\end{equation}
As we will see, this is always greater than one..


\section{Geodesic equations}
\label{sec4}


For the sake of completeness, we give below the geodesic equations in such a perturbed metric. For this aim, we need to consider
\begin{equation}
    ds^2=-\alpha^2(r,t)dt^2+\gamma_{11}(r,\theta,t)dr^2+\gamma_{22}(r,\theta,t)d\theta^2+\gamma_{33}(r,\theta,t)d\phi^2.
\end{equation}
From this it is easy to derive the Lagrangian
\begin{equation}
   L=-\left(-\alpha^2(r,t){\dot t}^2+\gamma_{11}(r,\theta,t)(\dot r)^2+\gamma_{22}(r,\theta,t){\dot\theta}^2+\gamma_{33}(r,\theta,t){\dot\phi}^2\right)^\frac{1}{2}
\end{equation}
%




where the dot means derivative with respect to the proper time $s$. Then, using the Euler-Lagrange equations one has
\begin{eqnarray}
    &&\frac{d}{ds}\left(\alpha^2{\dot t}\right)
    -\frac{1}{2}\frac{\partial\alpha^2}{\partial t}{\dot t}^2
    +\frac{1}{2}\frac{\partial\gamma_{11}(r,\theta,t)}{\partial t}{\dot r}^2
    +\frac{1}{2}\frac{\partial\gamma_{22}(r,\theta,t)}{\partial t}{\dot\theta}^2
    +\frac{1}{2}\frac{\partial\gamma_{33}(r,\theta,t)}{\partial t}{\dot\phi}^2=0
    \nonumber \\
    &&\frac{d}{ds}\left(\gamma_{11}{\dot r}\right)
    +\frac{\partial\alpha^2}{\partial r}{\dot t}^2
    -\frac{\partial\gamma_{11}(r,\theta,t)}{\partial r}{\dot r}^2
    -\frac{\partial\gamma_{22}(r,\theta,t)}{\partial r}{\dot\theta}^2
    -\frac{\partial\gamma_{33}(r,\theta,t)}{\partial r}{\dot\phi}^2=0
    \nonumber \\
    &&\frac{d}{ds}\left(\gamma_{22}{\dot\theta}\right)
    -\frac{\partial\gamma_{11}(r,\theta,t)}{\partial\theta}{\dot r}^2
    -\frac{\partial\gamma_{22}(r,\theta,t)}{\partial\theta}{\dot\theta}^2
    -\frac{\partial\gamma_{33}(r,\theta,t)}{\partial\theta}{\dot\phi}^2=0
    \nonumber \\
    &&\frac{d}{ds}\left(\gamma_{33}\dot\phi\right)
    =0.
\end{eqnarray}
Then, finally
\begin{eqnarray}
    \frac{d}{ds}[\alpha^2(r,t){\dot t}]
    -\frac{1}{2}\frac{\partial\alpha^2}{\partial t}{\dot t}^2
    +\frac{1}{2}\frac{\partial\gamma_{11}(r,\theta,t)}{\partial t}{\dot r}^2
   +\frac{1}{2}\frac{\partial\gamma_{22}(r,\theta,t)}{\partial t}{\dot\theta}^2
    +\frac{1}{2}\frac{\partial\gamma_{33}(r,\theta,t)}{\partial t}{\dot\phi}^2
    &=&0
    \nonumber \\
    \frac{d}{ds}[\gamma_{11}(r,\theta,t){\dot r}]
    +\frac{1}{2}\frac{\partial\alpha^2}{\partial r}{\dot t}^2
    -\frac{1}{2}\frac{\partial\gamma_{11}(r,\theta,t)}{\partial r}{\dot r}^2
   -\frac{1}{2}\frac{\partial\gamma_{22}(r,\theta,t)}{\partial r}{\dot\theta}^2
    -\frac{1}{2}\frac{\partial\gamma_{33}(r,\theta,t)}{\partial r}{\dot\phi}^2&=&0
    \nonumber \\
    \frac{d}{ds}[\gamma_{22}(r,\theta,t){\dot\theta}]
    -\frac{1}{2}\frac{\partial\gamma_{11}(r,\theta,t)}{\partial\theta}{\dot r}^2
   -\frac{1}{2}\frac{\partial\gamma_{22}(r,\theta,t)}{\partial\theta}{\dot\theta}^2
    -\frac{1}{2}\frac{\partial\gamma_{33}(r,\theta,t)}{\partial\theta}{\dot\phi}^2&=&0
    \nonumber \\
    \frac{d}{ds}[\gamma_{33}(r,\theta,t){\dot\phi}]&=&0
\end{eqnarray}

For a full radial motion we can set $\theta=\pi/2$ yielding
\begin{eqnarray}
    \frac{d}{ds}[\alpha^2(r,t){\dot t}]
    -\frac{1}{2}\frac{\partial\alpha^2}{\partial t}{\dot t}^2
    +\frac{1}{2}\frac{\partial\gamma_{11}(r,t)}{\partial t}{\dot r}^2
    +\frac{1}{2}\frac{\partial\gamma_{33}(r,t)}{\partial t}{\dot\phi}^2
    &=&0
    \nonumber \\
    \frac{d}{ds}[\gamma_{11}(r,t){\dot r}]
    +\frac{1}{2}\frac{\partial\alpha^2}{\partial r}{\dot t}^2
    -\frac{1}{2}\frac{\partial\gamma_{11}(r,t)}{\partial r}{\dot r}^2
    -\frac{1}{2}\frac{\partial\gamma_{33}(r,t)}{\partial r}{\dot\phi}^2&=&0\nonumber \\
    \frac{d}{ds}[\gamma_{33}(r,t){\dot\phi}]&=&0
\end{eqnarray}
The last equation of the set can be integrated out to give
\begin{equation}
    \dot\phi=\frac{A}{\gamma_{33}(r,t)}
\end{equation}
being $A$ an integration constant. This can be substituted in the other twos to give
\begin{eqnarray}
\label{eq:geods}
    \frac{d}{ds}[\alpha^2(r,t){\dot t}]
    -\frac{1}{2}\frac{\partial\alpha^2}{\partial t}{\dot t}^2
    +\frac{1}{2}\frac{\partial\gamma_{11}(r,t)}{\partial t}{\dot r}^2
    +\frac{1}{2}\frac{\partial\gamma_{33}(r,t)}{\partial t}\frac{A^2}{\gamma_{33}^2(r,t)}
    &=&0
    \nonumber \\
    \frac{d}{ds}[\gamma_{11}(r,t){\dot r}]
    +\frac{1}{2}\frac{\partial\alpha^2}{\partial r}{\dot t}^2
    -\frac{1}{2}\frac{\partial\gamma_{11}(r,t)}{\partial r}{\dot r}^2
    -\frac{1}{2}\frac{\partial\gamma_{33}(r,t)}{\partial r}\frac{A^2}{\gamma_{33}^2(r,t)}&=&0
\end{eqnarray}
Now, we know from eq.(\ref{eq:gammas}) that
\begin{eqnarray}
    \gamma_{11}(r,t)&=&U(r,t)\gamma_{11}^{(0)} \nonumber \\
    \gamma_{33}(r,t)&=&U(r,t)\gamma_{33}^{(0)}
\end{eqnarray}
and then
\begin{eqnarray}
    \frac{d}{ds}[\alpha^2(r,t){\dot t}]
    -\frac{1}{2}\frac{\partial\alpha^2}{\partial t}{\dot t}^2
    +\frac{1}{2}\frac{\partial U(r,t)}{\partial t}\left[{\dot r}^2\gamma_{11}^{(0)}
    +\frac{A^2}{U^2(r,t)\gamma_{33}^{(0)}}\right]
    &=&0
    \nonumber \\
    \frac{d}{ds}[\gamma_{11}(r,t){\dot r}]
    +\frac{1}{2}\frac{\partial\alpha^2}{\partial r}{\dot t}^2
    -\frac{1}{2}\frac{\partial U(r,t)}{\partial r}\left[{\dot r}^2\gamma_{11}^{(0)}
    +\frac{A^2}{U^2(r,t)\gamma_{33}^{(0)}}\right]&=&0.
\end{eqnarray}
This set can be solved only numerically. So, this approach does not lend itself for a straightforward computation of the radial velocity.

\section{Radial Velocity}
\label{sec5}

We consider a particle of mass $m$ moving in our metric. The definition of momenta is given by
\begin{equation}
    p_\alpha=g_{\alpha\beta}p^\beta.
\end{equation}
This yields the dispersion relation
\begin{equation}
    p_\alpha p^\alpha=-m^2.
\end{equation}
Similarly, we can derive the 4-velocity from this and is given by
\begin{equation}
    u_\alpha=(-\alpha^2{\dot t},\gamma_{11}{\dot r},\gamma_{22}{\dot\theta},\gamma_{33}{\dot\phi}).
\end{equation}
Then, the radial motion will be characterized by
\begin{equation}
\label{eq:vr}
    v_r=\frac{\gamma_{11}}{\alpha}\frac{dr}{dt}=\frac{U(r,\theta,t)}{\alpha(r,\theta,t)}\gamma_{11}^{(0)}\frac{dr}{dt}.
\end{equation}
One gets a {\em warp factor}, arising from the applied perturbation,
\begin{equation}
    U(r,t)=\exp\left[r_k\alpha_0^{-1}\int_0^tdt''\frac{\alpha(t'')}{\frac{3}{2}r_k\alpha_0^{-1}\int_0^{t''}\alpha(t')dt'+\sqrt{\gamma(0)}}\right]
\end{equation}
and we realize that, with this geometry, we can have an exponential growth of the radial velocity depending on the applied perturbation.

We can provide a closed form solution for a very simple case, a toy model. We take for a perturbation
\begin{equation}
    \alpha_1(t)=\frac{t}{\eta}
\end{equation}
being $\eta$ a constant. This is a linear time increasing term. Then,
\begin{equation}
    \alpha^2(r,t)=\alpha_0^2(r)+\frac{t}{\eta}.
\end{equation}
Then,
\begin{equation}
    U(r,t)=\exp\left[r_k\alpha_0^{-1}(r)\int_0^tdt''\frac{\sqrt{\alpha_0^2(r)+\frac{t''}{\eta}}}{\frac{3}{2}r_k\alpha_0^{-1}(r)\int_0^{t''}\sqrt{\alpha_0^2(r)+\frac{t'}{\eta}}dt'+\sqrt{\gamma(0)}}\right].
\end{equation}
This yields
\begin{equation}
    U(r,t)=\exp\left[r_k\alpha_0^{-1}(r)\int_0^tdt''\frac{\sqrt{\alpha_0^2(r)+\frac{t''}{\eta}}}{\frac{3}{2}r_k\alpha_0^{-1}(r)\eta
    \left[\frac{2}{3}\left(\alpha_0^2(r)+\frac{t''}{\eta}\right)^\frac{3}{2}-\frac{2}{3}\alpha_0(r)\right]+\sqrt{\gamma(0)}}\right].
\end{equation}
and
\begin{equation}
    U(r,t)=\exp\left[r_k\alpha_0^{-1}(r)
    \frac{3}{2}\eta
    \int_{\frac{2}{3}\alpha_0(r)}
    ^{\frac{2}{3}\left(\alpha_0^2(r)+\frac{t}{\eta}\right)^\frac{3}{2}}dx
    \frac{1}{\frac{3}{2}r_k\alpha_0^{-1}(r)\eta
    \left[x-\frac{2}{3}\alpha_0(r)\right]+\sqrt{\gamma(0)}}\right].
\end{equation}
Final result is
\begin{equation}
    U(r,t)=\frac{a\left(\alpha_0^2(r)+\frac{t}{\eta}\right)^\frac{3}{2}-b}{a\alpha_0^3(r)-b},
\end{equation}
being $a = r_k\eta\alpha_0^{-1}(r)$ and $b=r_k\eta-\sqrt{\gamma(0)}$. It is to see that this factor is always greater than one (this value is taken for $t=0$) and increasing as time increases.



From the formula for radial velocity we can derive the force. This will be obtained by the first derivative of eq.(\ref{eq:vr}). This yields
\begin{equation}
    \frac{dv_r}{d\tau}=\frac{d}{d\tau}\left[\gamma_{11}\frac{dr}{d\tau}\right].
\end{equation}
This gives, for a mass $M$,
\begin{equation}
   F=M\frac{dv_r}{d\tau}=M\frac{dt}{d\tau}\frac{d\gamma_{11}}{dt}\frac{dr}{d\tau}+\gamma_{11}\frac{d^2r}{d\tau^2}.
\end{equation}
This gives,
\begin{equation}
   F=M\frac{dv_r}{d\tau}=M\frac{1}{\alpha}\frac{d\gamma_{11}}{dt}\frac{dr}{d\tau}+\gamma_{11}\frac{d^2r}{d\tau^2}.
\end{equation}

In our toy model, we consider $\alpha_0\approx 1$ and $\gamma_{11}^{(0)}\approx 1$, so that
\begin{equation}
    \frac{d\gamma_{11}}{dt}=\frac{\frac{a}{\eta}\left(1+\frac{t}{\eta}\right)^\frac{1}{2}-\frac{db}{dt}}{a-b}+
    \frac{a\left(1+\frac{t}{\eta}\right)^\frac{3}{2}-b}{(a-b)^2}\frac{db}{dt}.
\end{equation}
This yields,
\begin{equation}
    \frac{d\gamma_{11}}{dt}\approx\frac{r_k\left(1+\frac{t}{\eta}\right)^\frac{1}{2}+2r\frac{dr}{dt}}{r^2}+
    \frac{r_k\eta\left(1+\frac{t}{\eta}\right)^\frac{3}{2}-r_k\eta+r^2}{r^4}2r\frac{dr}{dt}.
\end{equation}
Then, we get
\begin{eqnarray}
    F&\approx& M\left(1+\frac{t}{\eta}\right)^{-\frac{1}{2}}\left[
    \frac{r_k}{r^2}\left(1+\frac{t}{\eta}\right)^\frac{1}{2}+\frac{2}{r}\frac{dr}{dt}+
    \frac{2r_k\eta\left(1+\frac{t}{\eta}\right)^\frac{3}{2}-2r_k\eta+2r^2}{r^3}\frac{dr}{dt}\right]\frac{dr}{dt} \nonumber \\
    &&+
    M\frac{r_k\eta\left(1+\frac{t}{\eta}\right)^\frac{3}{2}-r_k\eta+r^2}{r^2}\frac{d^2r}{d\tau^2}.
\end{eqnarray}
with the simple kinematic law of motion $r(t)=r_0+v_0t$ it is easy to get
\begin{eqnarray}
    F&\approx& Mv_0\frac{r_k}{r^2(t)}+2M\left(1+\frac{t}{\eta}\right)^{-\frac{1}{2}}\frac{v_0^2}{r(t)}+ \nonumber \\
    &&M\left(1+\frac{t}{\eta}\right)^{-\frac{1}{2}}\frac{2r_k\eta\left(1+\frac{t}{\eta}\right)^\frac{3}{2}-2r_k\eta+2r^2(t)}{r^3(t)}v_0^2.
\end{eqnarray}
Force is non-null and dependent on the initial velocity and the sphere radius. It is interesting to note that the force tend to 0 as time increases but this corresponds to the unphysical case of a perturbation never turned off. This equation simplifies a lot if we can neglect the terms dependent on $r_k$. One has
\begin{equation}
    F\approx 3M\left(1+\frac{t}{\eta}\right)^{-\frac{1}{2}}\frac{v_0^2}{r(t)}.
\end{equation}
This result is independent on the sphere geometry or the Schwarzschild radius. Such a perturbation is not completely physical. So, we considered some others having the characteristic to be practically realizable. Considering the interior solution, for a perturbation like $\alpha_1=At^2$ we get
 \begin{figure}[H]
  \includegraphics[width=1\textwidth]{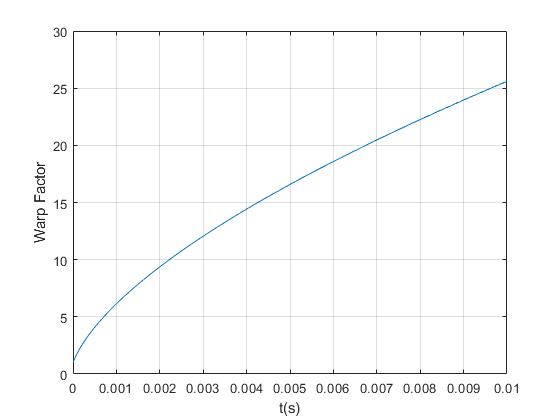}
  \caption{Warp factor for a $t^2$ perturbation with an equation of motion $r(t)=h_0+v_0 t$. \label{fig:pert1}},
\end{figure}
and for a sinusoidal perturbation  
\begin{figure}[H]
  \includegraphics[width=1\textwidth]{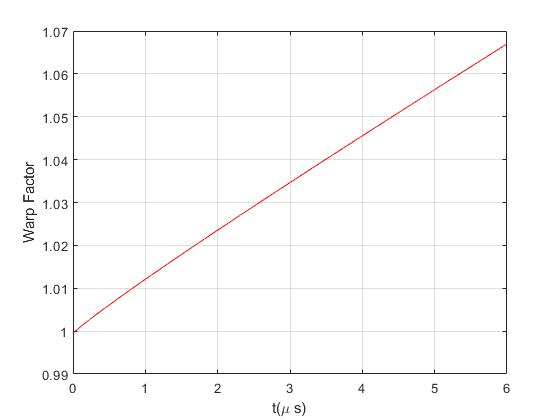}
  \caption{Warp factor for a $\sin(\omega t)$ perturbation with frequency 1 MHz and equation of motion $r(t)=h_0+v_0 t+kt^2$ . \label{fig:pert2}}
\end{figure}
As expected from the toy model, the warp factor is always greater than one and can reach significantly large values depending on the applied perturbation.

\section{Conclusions}
\label{sec6}

We have solved the Einstein equations for a strong perturbation in the case of a spherical symmetry solution. In this case, the perturbation series reduces to the case of a gradient expansion and the equations are amenable to an exact analytical treatment. We were able to show that, when a perturbation is properly applied, there appears a multiplicative warp factor on the radial velocity that can, in this way, increase exponentially in time. This warp effect does not require exotic energy and everything is completely in the realm of positive energy solutions of the Einstein equations, even if as a perturbation series.

We hope these results will find some application in the near future.

\end{document}